# Mapping a finite and an infinite Hadamard quantum walk onto a unique case of a random walk process

Arie Bar-Haim

Davidson Institute of Science Education, Weizmann Institute of Science, Israel

A new model that maps a quantum random walk described by a Hadamard operator to a particular case of a random walk is presented. The model is represented by a Markov chain with a stochastic matrix, i.e., all the transition rates are positive, although the Hadamard operator contains negative entries. Using a proper transformation that is applied to the random walk distribution after *n* steps, the probability distributions in space of the two quantum states $|1>$, $|0>$ are revealed. These show that a quantum walk can be entirely mapped to a particular case of a higher dimension of a random walk model. The random walk model and its equivalence to a Hadamard walk can be extended for other cases, such as a finite chain with two reflecting points

Various problems of quantum random walks have been investigated by many groups. For example, Aharonov et al. [1] explored quantum random walks, while Ambainis et al. [2–4] examined quantum walks on graphs. Bach et al. [5] investigated one-dimensional quantum walks with absorbing boundary conditions. Dür et al. [6] discussed quantum random walks in optical lattices. Moreover, Konno et al. [7] examined absorption problems and the eigenvalues of two-state quantum walks [8]. Mackay et al. [9] explored quantum walks in higher dimensions and Bartlet et al. [10] examined quantum topology identification in addition to various other problems [11,12].

Several studies have discussed the differences between random walks and quantum random walks, such as those investigated by Childs et al. [13] and Motes et al. [14] . The differences are manifested in various distribution functions that influence the moments of the dynamics

In the present study, a new model that maps a quantum random walk described by a Hadamard operator to a particular case of a random walk is presented.  The model is represented by a Markov chain with a stochastic matrix, i.e., all the transition rates are positive, although the Hadamard operator contains negative entries. Using a proper transformation that is applied to the random walk distribution after *n* steps, the probability distributions in space of the two quantum states |1>, |0>  are revealed. These show that a quantum walk can be entirely mapped to a particular case of a random walk model. The model also enables us to extend it to different boundary conditions, such as a reflecting point or a trap, as well as to obtain the distribution of each of the coin states separately, thereby gaining a better understanding of the entire system.

## Mapping a Hadamard operator to a four sites symmetric Markov chain

The discrete-time quantum random walk is defined by two operators [2,3]: the Hadamard operator that flips the quantum state of each site, and the shift operator, which moves the quantum states. The Hilbert space that governs the walk can be defined by a  tensor product of $\mathcal{HS} \otimes \mathcal{HC}$, where $\mathcal{HC}$ is a 2D Hilbert defined by the coin states, and $\mathcal{HS}$ is the space vector. The dynamic of one step can be determined by $u = S(I_d \otimes H)$, where $H$ is the Hadamard matrix, $I_d$ is a $d \times d$ unit matrix, and the subscript $d$ defines the space dimension. The tensor product is defined by a Kronecker product, and $S$ is the shift operator.

The Hadamard matrix, $H$, is defined by the following unitary operator:

$$H = \frac{1}{\sqrt{2}}\begin{bmatrix} 1 & 1 \\ 1 & -1 \end{bmatrix}, \qquad (1)$$

and the coin states can be defined by $|0>$ and $|1>$ as follows:

$$|0> = \begin{pmatrix} 1 \\ 0 \end{pmatrix} \qquad |1> = \begin{pmatrix} 0 \\ 1 \end{pmatrix}, \qquad (2)$$

The Hadamard operator can be mapped by four states of a Markov chain, as shown in Figure 1[15], with transition probabilities of $p = q = 0.5$.

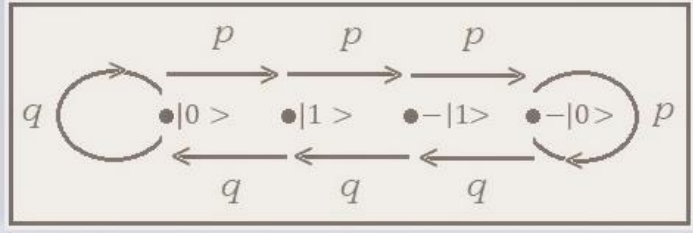

Figure 1. A Markov chain represents the transitions between four states. Note that each site is named with the following symbols respectively:

$$|0>, |1>, -|1>, -|0>$$

This can be shown explicitly by the following product:

$$H = \frac{1}{\sqrt{2}} BAB^T \qquad (3)$$

where $A$ is the transition probability matrix of the system described in Figure 1:

$$A = \begin{bmatrix} 0.5 & 0.5 & 0 & 0 \\ 0.5 & 0 & 0.5 & 0 \\ 0 & 0.5 & 0 & 0.5 \\ 0 & 0 & 0.5 & 0.5 \end{bmatrix}, \qquad (4)$$

and matrix $B$ is defined as

$$B = \begin{bmatrix} 1 & 0 & 0 & -1 \\ 0 & 1 & -1 & 0 \end{bmatrix}, \qquad (5)$$

which will be called in this context 'an interference matrix.'

Note that matrix $B$ interferes with the first and last entry of any column vector in the order of 4x1 and similarly interferes with the second and third entries. Also, note that $\text{Det}(H) \neq 0$, while $\text{Det}(A) = 0$.

## Building an RW model using the four-sites chain in the y-axis

The second operator, Hadamard walk, is the shift operator [2], which propagates the quantum states to the right or left, depending on whether the quantum states are $|0>$ or $|1>$. Combining the shift operator with the four-sites chain yields the RW model depicted in Figure 2. The horizontal direction, the x-axis, describes the movements due to the shift operator, and the vertical direction, the y-axis, represents the movements due to the Hadamard operator.

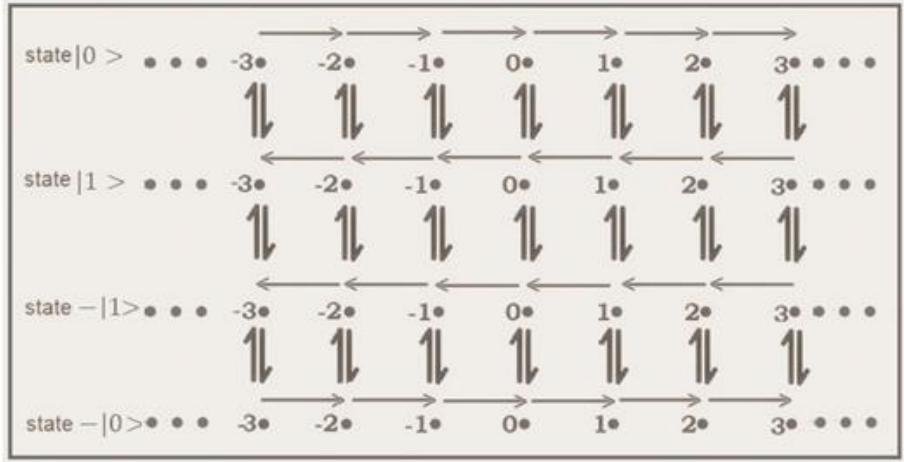

Figure 2. A two-step RW model. Note that the first to fourth rows are named as $|0>$, $|1>, -|1>, -|0>$

The main points of the RW model are as follows:

1. The shift operator is responsible for the movement in the horizontal direction (x-axis).
2. The transition matrix $A$ is responsible for the movement in the vertical direction (y-axis).
3. The movements in each direction occur one after the other, rather than simultaneously.

Table 1 describes a comparison between an infinite Hadamard walk and the RW model:

|  | Hadamard walk | Random walk model |
| --- | --- | --- |
| Hamdard Operator | $y = I_d \otimes H$ | $Y = I_d \otimes A$ |
| Shift operator | $x = Right \otimes Zero + Left \otimes One$ | $X = Right \otimes \hat{Z}ero + Left \otimes \hat{O}ne$ |
| The whole process | $u = xy=$ <br> $= Right \otimes ZeroH + Left \otimes OneH$ | $U = XY =$ <br> $Right \otimes \hat{Z}eroA + Left \otimes \hat{O}neA$ |

And the equivalence between the models is proved by the following relation[15]:

$$|u^n(I_d \otimes B)P(0)| = (\sqrt{2})^n |(I_d \otimes B)U^n P(0)| = (\sqrt{2})^n |(I_d \otimes B)P(n)| \qquad (6)$$

Where $n$ represents the step number,

$$Zero = \begin{bmatrix} 1 & 0 \\ 0 & 0 \end{bmatrix} \quad , One = \begin{bmatrix} 0 & 0 \\ 0 & 1 \end{bmatrix} \tag{7}$$

$$\hat{Zero} = \begin{bmatrix} 1 & 0 & 0 & 0 \\ 0 & 0 & 0 & 0 \\ 0 & 0 & 0 & 0 \\ 0 & 0 & 0 & 1 \end{bmatrix} , \hat{One} = \begin{bmatrix} 0 & 0 & 0 & 0 \\ 0 & 1 & 0 & 0 \\ 0 & 0 & 1 & 0 \\ 0 & 0 & 0 & 0 \end{bmatrix} \tag{8}$$

The *Right* and the *Left* matrices are $d \times d$ zero matrices, except those entries that appear above and beneath the main diagonal, respectively, and specifically $Right\ (j, j+1) = 1$ and $Left(j+1, j) = 1$ for any integer $j \in (1, d-1)$ as follows:

$$Right = \begin{bmatrix} 0 & 1 & 0 & 0 & 0 & \dots \\ 0 & 0 & 1 & 0 & 0 & \dots \\ 0 & 0 & 0 & 1 & 0 & \dots \\ 0 & 0 & 0 & 0 & 1 & \dots \\ 0 & 0 & 0 & 0 & 0 & \dots \\ \dots & \dots & \dots & \dots & \dots & \dots \end{bmatrix}, Left = \begin{bmatrix} 0 & 0 & 0 & 0 & 0 & \dots \\ 1 & 0 & 0 & 0 & 0 & \dots \\ 0 & 1 & 0 & 0 & 0 & \dots \\ 0 & 0 & 1 & 0 & 0 & \dots \\ 0 & 0 & 0 & 1 & 0 & \dots \\ \dots & \dots & \dots & \dots & \dots & \dots \end{bmatrix}, \tag{9}$$

$P(n)$ is a vectorization of the matrix that presents the population distribution of the RW model depicted in figure (2). The row vectors of this matrix are:

$$P_{|0>}(n) = [p_1(n), p_5(n), p_9(n), p_{13}(n) \dots p_{d-3}(n)] \tag{10}$$

$$P_{|1>}(n) = [p_2(n), p_6(n), p_{10}(n), p_{14}(n) \dots p_{d-2}(n)]$$

$$P_{-|1>}(n) = [p_3(n), p_7(n), p_{11}(n), p_{15}(n) \dots p_{d-1}(n)]$$

$$P_{-|0>}(n) = [p_4(n), p_8(n), p_{12}(n), p_{16}(n) \dots p_d(n)]$$

And stacking the columns of this matrix yields $P(n)$ as follows:

$$P(n) = [p_1(n), p_2(n), p_3(n), p_5(n), p_6(n), p_7(n) \dots \dots p_d(n)]^T \tag{11}$$

Applying $I_d \otimes B$ to $P(n)$ generates the following 2xd matrix:

$$\begin{bmatrix} P_{|0>}(n) - P_{-|0>}(n) \\ P_{|1>}(n) - P_{-|1>}(n) \end{bmatrix} \tag{12}$$

Thus, based on Eq. (6), the probability distribution of quantum states $|0>$ and $|1>$ after $n$ steps in space are, the following magntitude squared respectively:

$$\left|\psi_{|0>}(n)\right|^2 = 2^n \left|P_{|0>}(n) - P_{-|0>}(n)\right|^2 \tag{13}$$

$$\left|\psi_{|1>}(n)\right|^2 = 2^n \left|P_{|1>}(n) - P_{-|1>}(n)\right|^2$$

Figure 3 below presents known results of a Hadamard walk. The right figure describes probability distribution of Hamdard walk starting at the origin with a $|0>$ quantum after 100 steps. The calculations were done using Eq. (13) with initial condition of $P(0) = [0,0,0,0,0,0 \ldots 1,0,0,0,00 \ldots]^T$, where the 1 is the 401kth entry, in the case that $P(0)$ presents sites -100 till 100.

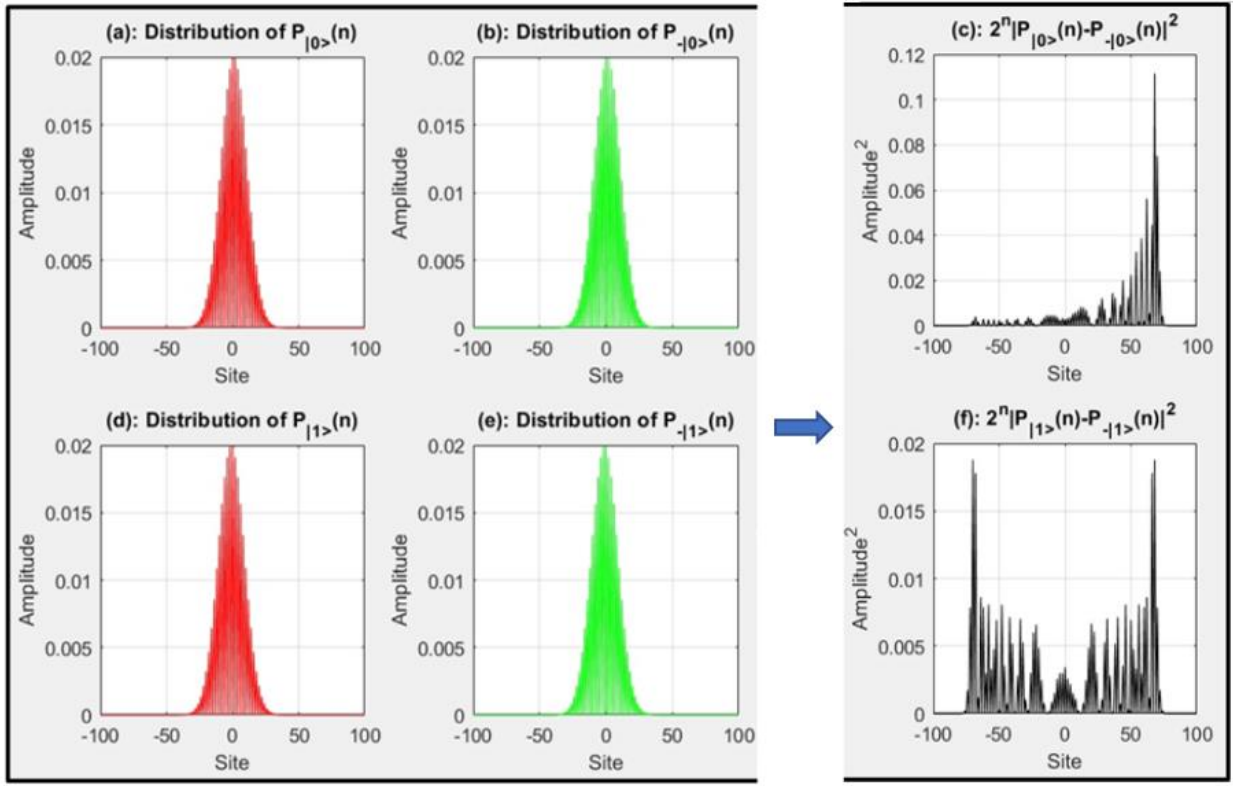

Figure 3. Graphs (a), (b), (d), and (e) present the RW distribution, based on it, graphs (c) and (f) present the probability distribution of the quantum states $|0>, |1>$.

The graphs in the first row describe:

(a) The RW distribution of state |0> in space after 100 steps, represented by $P_{|0>}(n)$

(b) The RW distribution of state -|0> in space after 100 steps, described by $P_{-|0>}(n)$

(c) The QRW probability distribution of the quantum state |0>, based on Eq. (13)

Similarly, the three graphs in the second row describe:

(d) The RW distribution of state |1> in space after 100 steps, described by $P_{|1>}(n)$

(e) The RW distribution of state –|1> in space after 100 steps, described by $P_{-|1>}(n)$

(f) The QRW probability distribution of the quantum state |1>, based on Eq. (13)

## Inserting boundaries into the RW model

There are two possibilities of reflecting barriers: switching between the states $|0>, |1>$ and $-|0>, -|1>$, or switching between the states $|0>, -|1>$ and $-|0>, |1>$, as presented correspondingly by the matrices $R_1, R_2$:

$$R_1 = \frac{1}{\sqrt{2}}\begin{bmatrix} 0 & 1 & 0 & 0 \\ 1 & 0 & 0 & 0 \\ 0 & 0 & 0 & 1 \\ 0 & 0 & 1 & 0 \end{bmatrix}, \quad R_2 = \frac{1}{\sqrt{2}}\begin{bmatrix} 0 & 0 & 1 & 0 \\ 0 & 0 & 0 & 1 \\ 1 & 0 & 0 & 0 \\ 0 & 1 & 0 & 0 \end{bmatrix} \tag{14}$$

The following figure describes a finite system with reflecting boundaries at locations -3 and 3, presented by $R_1$

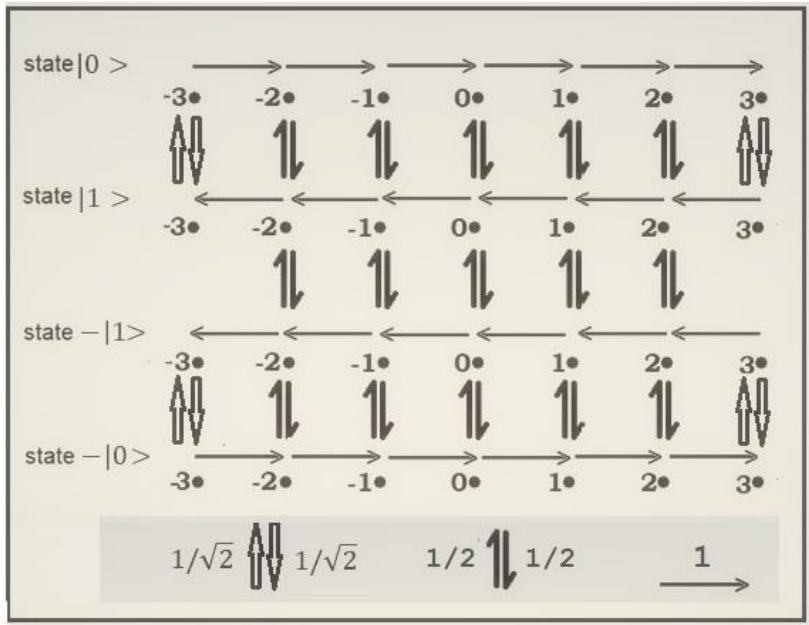

Figure 4. A finite two-step RW model. The horizontal direction describes the movements due to the shift operator; the vertical direction describes the movements due to the transition matrix $A$ and the boundaries of the system presented by $R_1$. At the bottom, there is a description of the transition probabilities values.

Note that the shift operator presented in the last figure is not unitary. To make it unitary, it is necessary to add the following connection as depicted in Figure 5, i.e., to make the system cyclic.

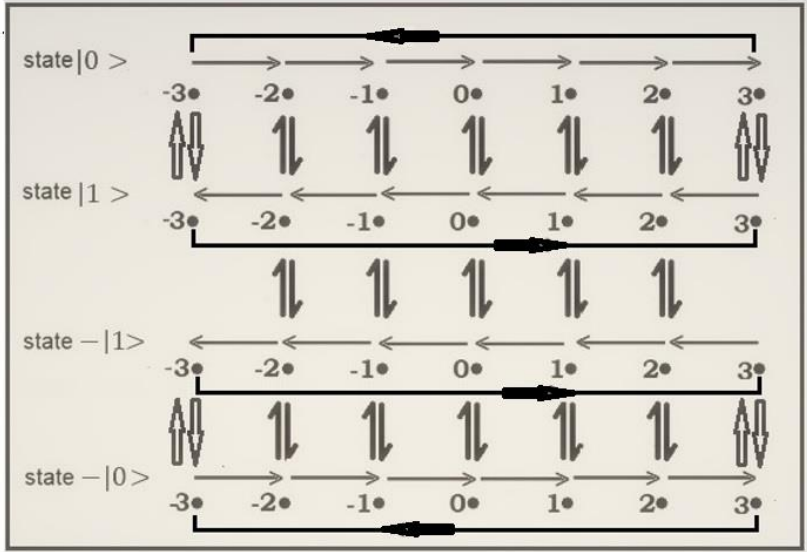

Figure 5. Cyclization of the x-axis operator

Notably, this cyclization is unnecessary, since actually no population propagates from the left side of the system to the other side of the system for any case of initializing anywhere other than the boundaries. This can be explained as follows:

Assume that at step $n$, the one states $|1>$ reach site -3, and then the y-axis ($R_1$) operates and therefore switches this state into the zero states $|0>$. Then the shift operator operates again, but none of the ones, $|1>$, remain on the left boundary. Therefore, nothing propagates to the other side of the system. This occurs simialrly at the right side, for the $|0>$ states.

The following describes a comparison between a finite Hadamard walk and the RW

Table a comparison between a finite Hadamard walk and the RW model

|  | Finite Hadamard walk | Finite random walk |
| --- | --- | --- |
| Hadamard operator | $y_{rp} = (I_d - Z) \otimes H + Z \otimes r_1$ | $Y_{rp} = (I_d - Z) \otimes A + Z \otimes R_1$ |
| Shift operator | $x = Right \otimes Zero + Left \otimes One$ | $X = Right \otimes \hat{Z}ero + Left \otimes \hat{O}ne$ |
| The whole process | $u_{rp} = xy_{rp}$ | $U_{rp} = XY_{rp}$ |

The equivalence between the models is proved by the following relation[15]:

$$|u_{rp}(I_d \otimes B)| = \sqrt{2}|(I_d \otimes B)U_{rp}|$$

where

$$r_1 = \begin{bmatrix} 0 & 1 \\ 1 & 0 \end{bmatrix} \qquad (15)$$

and the Z matrix is a $d \times d$ zero matrix, except for two entries at the beginning and the end of the main diagonal, namely: $Z(1,1) = 1; Z(d,d) = 1$.

Figure 6 describes the probability distribution of a quantum random walk on a finite chain of 25 sites after 35 and 65 steps, presented by the random walk model in Figure 4 with 25 sites, where each site, except the boundaries, initializes at $(|0> - |1>)/\sqrt{46}$.
Note that the system initializes with equally distributed states |0> and –|1>, then switches after 35 steps, mostly to state |0>, and after 65 steps mostly goes back to state |1>.

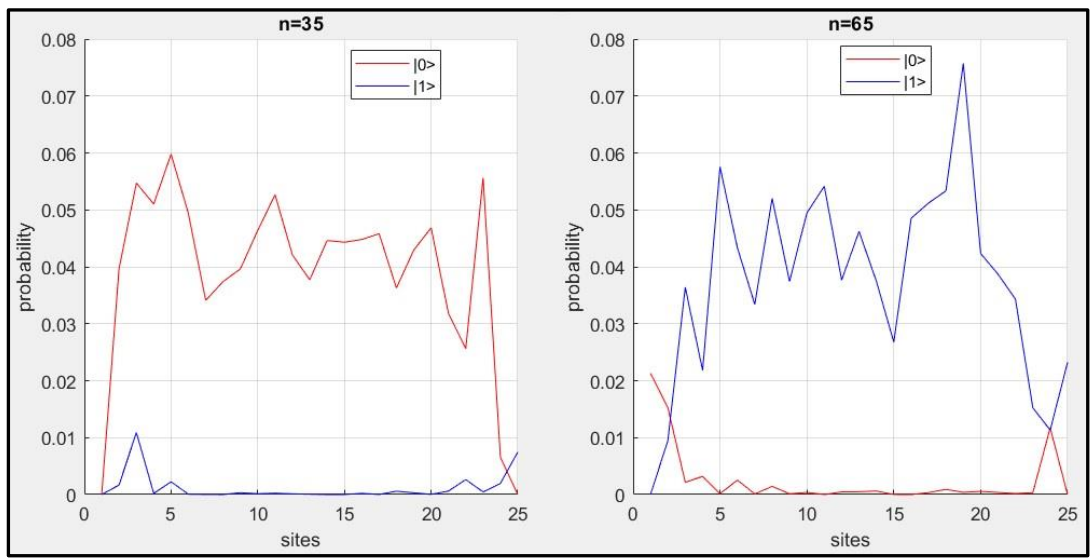

## Summary

A new model that maps a quantum random walk to a particular case of a RW model was presented here for both finite and infinite cases. The RW walk model has completely different properties. For example, an infinite system has a stochastic matrix; namely, the system conserves the random walk population. In addition, when using the transformation introduced here, the sum of the square amplitude of the quantum states is preserved as a unitary transformation.
Above all, this model shows that the dynamics between the RW model and a QRW are very similar, and it is the interference at the end of the process that makes the results of the QRW so different from a RW.